\documentstyle{article}

\font\tenbg=cmmib10 at 10pt

\def \rvecphi{{\hbox{\tenbg\char'036}}}

\begin{document}
{\Large
  {\centerline {\bf{A Cosmic Battery Reconsidered}}
}

\bigskip
\normalsize
{\centerline {\bf G.S. Bisnovatyi-Kogan,}}

{\it Space Research Institute, Russian Academy of Sciences,
Moscow, Russia; {\centerline{gkogan@mx.iki.rssi.ru}}}

\medskip

{\centerline {\bf R.V.E. Lovelace}}

{\it Department of Astronomy, Cornell University, Ithaca, NY
14853-6801;\\ {\centerline {rvl1@cornell.edu}}}

\medskip

{\centerline {\bf V.A. Belinski}}

{\it National Institute of Nuclear physics (INFN) and
International Center of Relativistic Astrophysics (ICRA), Rome,
Italy\\
 {\centerline{and}}\\
 Institute des Hautes Etudes
Scientifiques (IHES), Bures-sur-Yvette, France;\\
{\centerline{belinski@icra.it}}}

\begin{abstract}

   We revisit the problem of magnetic field
generation in accretion flows onto black
holes owing to the excess radiation force on
electrons.  This excess force may
arise from  the Poynting-Robertson effect.
    Instead of a recent claim of the
generation of dynamically important
magnetic fields, we establish
the validity of earlier results from 1977
which show only small magnetic fields are
generated.
   The radiative force
causes the magnetic field to initially grow linearly
with time.
However, this linear growth holds
for only a {\it restricted} time interval which is of
the order of the accretion time of the matter.
    The large magnetic fields recently found
result from the fact that the
linear growth is unrestricted.
    A  model of the Poynting-Robertson
magnetic field generation close to the horizon of
a Schwarzschild black hole is solved
exactly using General Relativity, and  the field is also
found to be dynamically insignificant.
    These weak magnetic fields may however be important as seed
fields for dynamos.

\end{abstract}

keywords {\it accretion, accretion disks---galaxies:
active---plasmas---magnetic fields---stars: magnetic
fields---X-rays: stars}

\section{Introduction}

    The classical battery
mechanism of  magnetic field generation is
connected with a noncoincidence of surfaces of
constant pressure and constant density, where
forces connected with pressure gradients become
nonpotential or rotational.
    In this situation no static
equilibrium in the gravitational field is
possible.
    When considering separately the  motion of
electrons and ions, there is always  a difference
in the velocities of electrons and ions which creates
electric currents and an associated magnetic
field.
     Self-induction is very important in the
battery mechanism, determining the rate of increase of
the magnetic field.

     Along with  ion and electron pressure
gradients, a nonpotential force field may
arise due to the radiation force which
acts predominantly  on the electrons.
     In a spherically symmetric star, the
radiation force has a potential so that
no magnetic field is  generated: equilibrium in the
two-fluid plasma results from a
  distribution of an electric charge and a
static radial electric field.
    For {\it geometrically-thin, optically-thick}
accretion disks, Bisnovatyi-Kogan and
Blinnikov (1977;  hereafter BKB) showed that the
radiation force above the disk has a nonpotential or
rotational component.
    Under this condition, no electric
charge distribution can give a static
equilibrium.
    Instead, electric
currents and a corresponding magnetic field are
generated.
    The radiation forces above a thin disk
gives rise to poloidal electrical current flow and
a toroidal magnetic field.

    In accretion flows at very low mass accretion
rates,  an {\it op\-ti\-cal\-ly-thin}, {\it
ge\-o\-met\-ri\-cal\-ly-thick} accretion flow is possible
(Shapiro, Lightman, \& Eardley, 1976) where the ion temperature is
close to the virial  temperature.
    In the absence of a magnetic field, and
neglecting  relaxation processes
between electrons and ions except for binary
collisions
these flows are referred to as
advection dominated accretion flows (ADAF)
(Ichimaru 1977; Narayan \& Yi 1995).
    In the ADAF regime the radiative efficiency of
accretion may be very low, $\sim 10^3$  times
less then the standard value for a geometrically
thin, optically thick accretion disk. Account of
processes connected with the presence of a magnetic
field increases the efficiency  up to
at least $1/3$
of the standard value (Bisnovatyi-Kogan \&
Lovelace 1997, 2000). Nevertheless, the disk
remains  geometrically thick in the
optically thin regime due to high ion temperature.

   Contopoulos and Kazanas (1998; hereafter
CK) proposed that
a cosmic battery may operate in
ADAF accretion flows owing to the
Poynting-Robertson (PR) effect.
    The PR effect acts to generate a
toroidal electrical current and poloidal magnetic field.
    The authors found that the
magnetic field may be amplified up to $\sim 10^7$G
in the vicinity of a black hole of stellar mass.
    Note that the PR
mechanism of magnetic field generation is similar to the mechanism
of BKB based on the nonpotential radiative force, where the
magnetic field reached values of $\leq 10$G for a stellar mass
black hole.
   In  an optically thin disk both mechanisms
act together leading to the generation of toroidal
and poloidal components of the magnetic field.
    The influence of the PR effect on the dynamics
of the surface layer of an accretion disk was
treated by Mott and Lovelace (1999).

    Here, we analyze the difference in conclusions
between CK and BKB.
   The radiative force initially
causes the magnetic field to grow linearly
with time.
However, this linear growth holds
for only a {\it restricted} time interval which is of
the order of the accretion time of the matter.
In CK the interval of linear growth is unrestricted.
     Even though we conclude that the magnetic field
due to the radiation force is weak,
it may have a role as a seed field
for an $\alpha-\omega$ dynamo (see for example
Colgate and Li 2000).

    Section 2 treats the generation of toroidal
field for the case of a thin disk, while \S 3
treats the generation of poloidal field in
an ADAF flow.
     Section 4 gives a General
Relativistic treatment of a simplified model of
PR magnetic field generation in an accretion flow close to a
Schwarzschild black hole.
   An appendix  gives an explicit solution to the
non-relativistic induction
equation for the magnetic field generated by
the PR effect in an ADAF flow.

\section{Radiatively induced current and toroidal\\
magnetic field production in accretion disks}

   Above a geometrically thin accretion disk around a
black hole, the electrons are acted on by a
nonpotential radiation force ${\bf F}_L$ due to Thomson
scattering.  This was calculated by BKB,
\begin{equation}
\label{eq1}
{\bf F}_L=-R\cos \theta~ {\bf \nabla}
\phi_L=(F_{LR},\,F_{L\theta},\,0)~,
\end{equation}
where a spherical coordinate system
$(R,\theta, \phi)$ is used, and
$\phi_L$  is
the ``pseudopotential'' of the radiation
force, which may be  expressed as
\begin{equation}
\label{eq2}
\phi_L=\frac{\sigma_T}{c}
\int_{r_{in}}^{\infty}\frac{H(r)\, r\,dr}
{(R^4+r^4+2R^2r^2\cos 2\theta)^{1/2}}~.
\end{equation}
Here, the disk thickness is neglected
and the cylindrical radius is $r=R\cos\theta$.
      The function $H(r)$ is the radiative
flux emitted per unit area from one side of the disk.
     In the standard local accretion model,
\begin{equation}
\label{eq3}
H(r)=\frac{3}{8\pi}\frac{GM \dot
M}{r^3}{\cal J}~,
\end{equation}
where ${\cal J} \equiv 1-(r_{in}/r)^{1/2}$, and
$r_{in}=3r_S = 6GM/c^2$ is the
inner radius of the disk for a non-rotating
black hole   of Schwarzschild
radius  $r_S$.
      In the disk plane, $\theta=\pi/2$,
the radiative force is perpendicular
to the disk,
\begin{equation}
\label{eq4}
F_{L\theta}=-\frac{\sigma_T}{c}H(r)
=\frac{3GMm_p}{r^2}{r_{in}
\over r}\frac{L}{L_{Edd}}{\cal J}~,
\end{equation}
where
\begin{equation}
\label{eq5} L= \frac{GM{\dot M}}{2r_{in}},\qquad
L_{Edd}=\frac{4\pi c GM m_p}{\sigma_T}~,
\end{equation}
where $\sigma_T$ is the Thomson cross section.
Due to interaction of the
radiation flux mainly with the electrons, the
accretion disk becomes  positively charged up to
a value where the electrostatic  attraction of
the   electrons balances the radiation force.
    The
vertical component of the electrical field
strength $E_{\theta}$ in the disk plane is
written as
\begin{equation}
\label{eq6}
E_{\theta}(r)=-\frac{\sigma_T}{c|e|}
H(r)=-E_0\frac{L}{L_{Edd}}
\left(\frac{r_{in}}{r}\right)^3
{\cal J}~,
\end{equation}
where
\begin{equation}
\label{eq7}
E_0(M) \equiv \frac{m_pc^4}{12|e|GM}\approx
1.76 \frac{M_{\odot}}{M}~ {\rm cgs}
\approx 528\frac{M_{\odot}}{M}{{\rm V}\over {\rm cm}}
\end{equation}
Thus, the surface charge-density of the
disk is
$$
\Sigma_e(r)=\frac{E_{\theta}(r)}{2\pi}~.
$$
The influence of this charge on
the structure and stability of the accretion disk
is negligible.

Both the gravitational and electrical forces have a
potential, so that they cannot balance the nonpotential
radiation force.
   Due to the
radiation and electric forces, electrons move
with respect to protons, which to a first
approximation do not acquire the poloidal motion.
     Thus a poloidal electrical current is
generated with an associated
toroidal magnetic field.
   The finite
disk thickness may create  poloidal motion of
all of the matter
of the accretion disk,
similar to meridional circulation in
rotating stars (Kippenhahn \& Thomas 1982).
     The absence of
this circulation occurs for a unique dependence of the
rotational velocity over the disk thickness,
$\Omega(z)$.

To estimate the magnetic field strength, we
write the electromotive force (EMF) as
\begin{equation}
\label{eq8}
{\cal E}= {1\over e}\oint {\bf F}_L \times
d{\bf l}=
{1\over e}\int\int d{\bf S}\cdot{\bf \nabla \times
F}_L
\sim E_{\theta} h~,
\end{equation}
where $h$ is the half-thickness of the disk.
     Thus the stationary  current-density is
\begin{equation}
\label{eq9}
J_{st} \sim \frac{\sigma_e {\cal
E}}{r}
\sim \frac{\sigma_e E_{\theta}h}{r}~,
\end{equation}
where $\sigma_e$ is the conductivity
of the disk plasma.
    The stationary toroidal magnetic
field (BKB) is
\begin{equation}
\label{eq10} B_{\phi 0} \sim \frac{4\pi}{c} J_{st}h
\sim \frac{4\pi \sigma_e}{c}
\frac{E_{\theta}h^2}{r}~.
\end{equation}
In the radiation-dominated inner region of
the standard $\alpha$-disk model, $h$ can be written as
\begin{equation}
\label{eq11} h=3\frac{L}{L_{Edd}}
{\cal J}~r_{in}~,
\end{equation}
(Shakura 1972; Shakura \& Sunyaev 1973).
    Finally, we obtain the stationary
value of toroidal magnetic field in the disk,
$$
\label{eq12}
B_{\phi 0} \sim
\frac{36\pi\sigma_e}{c}E_0
\left(\frac{L}{L_{Edd}}\right)^3
\left(\frac{r_{in}}{r}\right)^4 r_{in}{\cal J}^3~,
$$
\begin{equation}
=\frac{12\pi\sigma_e h}{c}E_0
\left(\frac{L}{L_{Edd}}\right)^2
\left(\frac{r_{in}}{r}\right)^4 {\cal J}^2~.
\end{equation}
We next discuss the value of conductivity $\sigma_e$.

Bisnovatyi-Kogan and Blinnikov
(1977)  considered different
values of the plasma  conductivity $\sigma_e$,
namely, the conductivity owing to binary
collisions,
$\sigma_{Coul}$, and the effective
conductivity $\sigma_{eff}$ derived by
Vainshtein (1971),
\begin{equation}
\label{eq13}
\sigma_{Coul}\approx 3\times 10^6 T^{3/2} s^{-1},
\quad
\sigma_{eff}=
{\sigma_{coul}
\over \sqrt{Re_{m0}}}~.
\end{equation}
The magnetic Reynolds number
in a turbulent plasma
is defined as
\begin{equation}
\label{eq14}
  Re_{m0} \equiv \frac{4\pi \sigma_{coul} v_t
h}{c^2}~,
\end{equation}
where the turbulent velocity $v_t$
in an $\alpha$-disk is $v_t=\alpha c_s$,
where $c_s=\sqrt{p/\rho}$.

   In addition to the values of equation (13),
we consider the conductivity in the
presence of  well-developed turbulence
(Bisnovaty-Kogan \& Ruzmaikin 1976),
\begin{equation}
\label{eq14a}
\sigma_{turb}=\frac{c^2}{4\pi \alpha h c_s}
=\frac{\sigma_{coul}}{Re_{m0}}~.
\end{equation}

   Estimations of the stationary field $B_{\phi0}$
of equation (12)
in the radiation dominated inner region
of the disk around a stellar mass
black hole gives values (BKB)
for two cases in (\ref{eq13}) of $B_{\phi 0}\sim
10^{13}$G for $\sigma_e=\sigma_{Coul}$
and $B_{\phi 0}\sim 10^8$G
for $\sigma_e=\sigma_{eff}$,
with $Re_{m0}
\sim 3\times 10^{10}$.
    For a turbulent conductivity
$\sigma_e=\sigma_{turb}$, we obtain
$B_{\phi 0} \sim 10\,$G for
$\alpha=0.1$.

   The time-scale $\tau_m$ for reaching the stationary
field given by equation (12)  is
determined by the self-induction of the disk.
   This is equivalent to the ``${\cal L}$ over ${\cal R}$
time'' of circuit with inductance ${\cal L}$ and
resistance ${\cal R}$.
    This time-scale is equal to
\begin{equation}
\label{eq15}
\tau_m \simeq \frac{4\pi \sigma_e h r}{c^2}~.
\end{equation}
     The crossing time of matter passing
through the radiation dominated region
of the disk is
\begin{equation}
\label{eq16} t_c \approx \frac{r}{v_r}~.
\end{equation}
     During the time $t_c$
there is  linear growth of the magnetic field
after which the matter falls into the black hole.
    Thus the stationary value of the large
scale toroidal magnetic field is
\begin{equation}
\label{eq17}
B_{\phi} \approx B_{\phi
0}\frac{t_c}{\tau_m}=3\frac{c}{v_r}
E_0\left(\frac{L}{L_{Edd}}\right)^2
\left(\frac{r_{in}}{r}\right)^4{\cal J}^2~.
\end{equation}

     For the case of a turbulent
conductivity $\sigma_e=\sigma_{turb}$, the growth
time-scale of  the magnetic field, using equations
  (\ref{eq14a}) and
(\ref{eq15}), is equal to
\begin{equation}
\label{eq18}
\tau_{m}^{turb} \approx \frac{r}{\alpha c_s}~.
\end{equation}
Taking into account that
  $v_r = \alpha c_s
(L/L_{Edd})(r_{in}/{r})<\alpha c_s$ and $t_m^{turb}<t_c$, we find
\begin{equation}
\label{eq20}
B_{\phi}=B_{\phi0}^{turb} \approx 3\frac{c}{\alpha c_s}
E_0\left(\frac{L}{L_{Edd}}\right)
\left(\frac{r_{in}}{r}\right)^3{\cal J}^2~.
\end{equation}
In that
$$
c_s=7\times 10^9~ \frac{\rm cm}{\rm
s}\,\,
\left(\frac{L}{L_{Edd}}\right)
\left(\frac{r_{in}}{r}\right)^{3/2}{\cal J}~.
$$
The strength of
the stationary toroidal magnetic field
produced by the battery effect in the
radiation-dominated region of an accretion disk
with the turbulent or higher conductivity from (\ref{eq13}) is
equal to
\begin{equation}
\label{eq21}
B_{\phi}\approx\frac{22}{\alpha}
\left(\frac{M_{\odot}}{M}\right)
\left(\frac{r_{in}}{r}\right)^{3/2}{\cal J}.
\end{equation}
At a distance $r=3r_{in}$, we have
\begin{equation}
B_{\phi}\approx
\frac{2}{\alpha}\left(\frac{M_{\odot}}{M}\right)~{\rm
G}~.
\end{equation}
This agrees with the findings of Bisnovatyi-Kogan and
Blinnikov (1977).
    The corresponding magnetic energy-density
is very much less than the energy-density
associated with the turbulent motion in
the disk $\rho v_t^2/2$.

\section{Production of a poloidal magnetic field
in  optically thin accretion flows by\\ Poynting-Robertson effect}

    In  optically thin accretion flows (Shapiro,
Lightman, \& Eardely 1976;
Ichimaru 1977; Narayan \& Yi 1995), the
radiation flux interacts with the
inspiraling matter by
the Poynting-Robertson (PR) effect (Robertson 1937).
   Analysis by Shakura (1972) showed that
the PR effect was negligible for optically
thick accretion disks.
   Contopoulos and Kazanas (1998) studied
the PR effect as a mechanism for
generating poloidal
magnetic field in an optically thin accretion
flow.
      They concluded that dynamically important
magnetic field strengths could result
from this effect.
   Here, we reconsider the PR effect for quasi-spherical
advection dominated accretion flows (ADAF, Narayan
\& Yi 1995).

    The linear growth of
the magnetic field due to the radiative
force on the electrons found by CK is
similar to that analyzed by
Bisnovatyi-Kogan and Blinnikov (1977), but the PR
effect implies an additional (small) numerator,
$(v_{\phi}/c)$.
     Also, for a  quasi-spherical accretion flow the
characteristic scale is $r$ instead of $h$, and the
quasi-spherical luminosity is $L/(4\pi r^2)$ instead of
$H$ in (\ref{eq3}).
    Then, using equations (\ref{eq10}),
(\ref{eq15}), (\ref{eq17}), we obtain the
rate of growth the poloidal magnetic field
due to the PR effect, which is equivalent to
the expression obtained by CK,
\begin{equation}
\label{eq25}
B_{z} \approx B_{z
0}\frac{t}{\tau_m}=\frac{E_0}{3\alpha}
\left(\frac{L}{L_{Edd}}\right)
\left(\frac{r_{in}}{r}\right)^2
\left(\frac{tv_r}{r}\right).
\end{equation}
Here, $E_0$ is defined in equation
(\ref{eq6}), and $r_{in}$ in
equation (\ref{eq3}).

    Now it is essential to take into account, that an
element of matter with the induced magnetic
field reaches the black hole in time
$t_c \approx r/v_r$.
(Damping of the magnetic field
as it approaches the black hole horizon is discussed
in Appendix A.)
This means that the magnetic field grows {\it only}
during the time $t_c$.
    Consequently, the magnetic field reaches
a stationary value
\begin{equation}
\label{eq26}
B_{z} \approx \frac{E_0}{3\alpha}
\frac{L}{L_{Edd}}
\left(\frac{r_{in}}{r}\right)^2
\approx \frac{0.7}{\alpha}
\frac{L}{L_{Edd}}
\frac{M_{\odot}}{M}
\left(\frac{r_{in}}{r}\right)^2
\end{equation}
   Taking into account that
 the luminosity is $\leq 10^{-3}
L_{Edd}$ for optically thin accretion, and taking $\alpha=0.1$, we
get a maximum value of the magnetic field created by the PR effect
in an ADAF to a black hole,
\begin{equation}
\label{eq27}
B_{z} \approx 7 \times 10^{-3}
\left(\frac{M_{\odot}}{M}\right)~{\rm G}~.
\end{equation}
    This estimate of the field is about
10 orders of magnitude less that the value
obtained by CK.
     The difference in the estimates results from
the fact that CK assume that magnetic flux
accumulates continuously near the black hole
during a long time, reaching the equipartition
with the kinetic energy.
    The accumulation actually occurs only
during the time  the plasma
(which carries the field  or current loops)
takes to move inward to the black hole horizon
(Bisnovatyi-Kogan and Ruzmaikin, 1976).
    The current
loops created by the PR effect
  disappear as  the matter
approaches the horizon (see Appendix A).
    In the case
of accretion onto a neutron star or a white dwarf,
matter containing the current loops
merges with the stellar matter, which is
typically much more strongly magnetized.
    After merging, the matter
becomes optically thick,
the action of PR effect  stops,
  penetration of matter
into the magnetosphere of the star occurs, and
interaction of the accretion flux with the
stellar surface takes place.

\section{
Magnetic field Generation in the vicinity of the
Schwarzschild radius}

     Here we consider  the Poynting-Robertson
magnetic field generation on the accretion
flow near a Schwarzschild black hole.
    The accreting matter is assumed to
radially towards the black hole with
velocity $v_r$.
    As before we consider the case  of
high conductivity matter where
$t_m \gg t_c$.
    In a non-accreting flow, the magnetic field
can grow linearly with time as
accepted by CK.
       As  mentioned above, there is
an important relativistic effect close to the black hole:
    Current
loops in the accreting matter which approach the
horizon of a black hole cannot
produce a magnetic field visible
by an external observer.
    This effect is
related  to the damping of the
magnetic field in a collapsing star (Ginzburg \& Ozernoi
1964).

The azimuthal force and the corresponding
azimuthal electric field
due to the Poynting-Robertson effect are
\begin{equation}
\label{a1}
F_{PR}=\frac{L\sigma_T}{4\pi c r^2} \frac{v_{\phi 0}}{c}
\sin{\theta}~,\quad
E_{\phi}^{(ph)}=\frac{1}{{|e|}} F_\phi^{PR}~, \quad
\end{equation}
where
\begin{equation}
\label{a1.1}
v_{\phi 0}=A\sqrt{\frac{GM}{r}}~,\quad
A\sim 1~,
\end{equation}
and where the $(ph)$ superscript indicates the
physical value of the field component.
    Note that in a strictly spherical
accretion flow there is no
azimuthal EMF.
   However, in the approximate model considered here
the infalling matter has a small rotation so that
the PR force is small and these have radial inflow
only slightly.

We assume a Schwarzschild metric,
\begin{equation}
\label{a2}
ds^2=g_{00}c^2 dt^2 +g_{11}dr^2 +g_{22}d\theta^2+g_{33}d\phi^2~,
\end{equation}
where
$$
g_{00}=\left(1-\frac{r_g}{r}\right)~,\quad
g_{11}=-\left(1-\frac{r_g}{r}\right)^{-1}~,
$$
$$
g_{22}=-r^2~,\quad
g_{33}=-r^2 \sin^2\theta~,
$$
and $r_g \equiv 2GM/c^2$.
The matter is free-falling in the radial direction
with the nonzero components of
a $4-$velocity
\begin{equation}
\label{a3}
u^0=\left(1-\frac{r_g}{r}\right)^{-1}~,\quad u_0=1,
\end{equation}
$$
u^r=-\sqrt{\frac{r_g}{r}}~,\quad
u_r=\sqrt{\frac{r_g}{r}}\left(1-\frac{r_g}{r}\right)^{-1}~.
$$

In any $4-$dimensional time-space
with metric $g_{ik}$ (Latin indices
takes values $0,~1,~2,~3$),
the electric $E_{\alpha}$ and magnetic
$B^{\alpha}$ fields
(Greek indices run over the values $1,~2,~3$)
in $3-$space are defined through the
antisymmetric  electromagnetic
field tensor $F_{ik}=-F_{ki}$,
with zero diagonal components(Landau \& Lifshitz 1988),
\begin{equation}
\label{a11}
B^{\alpha}=-\frac{1}{2\sqrt{\gamma}}
\varepsilon^{\alpha\beta\gamma}F_{\beta\gamma}~,
\quad E_{\alpha}=F_{0\alpha}~,
\end{equation}
and
$$
F_{\alpha\beta}=-\sqrt{\gamma}
\varepsilon_{\alpha\beta\gamma} B^{\gamma}~,
$$
where $\varepsilon_{\alpha\beta\gamma} \equiv
\varepsilon^{\alpha\beta\gamma}$
is the usual antisymmetric $3-$dimensional  symbol
$(\varepsilon_{123}=1)$,
and $\gamma$ is the determinant of
the $3-$dimensional metric tensor,
obtained by splitting
the metric tensor $g_{ik}$ into
space ($\gamma_{\alpha\beta}$) and
time ($h$) parts as
\begin{equation}
\label{a6}
\gamma_{\alpha\beta}=-g_{\alpha\beta}+
g_{0\alpha}g_{0\beta}/g_{00}~,\quad
h=g_{00}~.
\end{equation}
  For the
Schwarzschild metric (\ref{a2}), $\gamma_{\alpha\beta}=-g_{\alpha\beta}$.
      The first pair of Maxwell equations can be written as
$$
F_{ik;l}+F_{li;k}+F_{kl;i}=0
$$
or
\begin{equation}
\label{a13a}
\frac{1}{c}\frac{\partial(\sqrt{\gamma}B^{\alpha})}{\partial t}+
\varepsilon^{\alpha\beta\gamma}\frac{\partial E_{\gamma}}{\partial
x^{\beta}}=0, \,\,\,\frac{\partial(\sqrt{\gamma}B^{\alpha}
)}{\partial x^{\alpha}}=0~.
\end{equation}
Note that the  physical $r$ and $\theta$ components of the magnetic
field  in this reference
frame are  $\sqrt{-g_{11}}B^r$ and $\sqrt{-g_{22}}B^\theta$
respectively.
Thus the dimensions of $B^\theta$
are ${\rm length}\times B^r$.

    In a perfectly conducting medium moving
with $4-$velocity $u^i$, we have  $F_{ik}u^k=0$,
which corresponds to a vanishing electric
field in the comoving frame.
    This gives
\begin{equation}
\label{a12.1}
  E_{\alpha}=-\sqrt{-g}\,\,\varepsilon_{\alpha\beta\gamma}
({v^{\beta}}/{c}) B^{\gamma}~.
\end{equation}
In the presence of an externally imposed electric
field $E_\alpha^{ext}$,
the electrical field
$E_{\alpha}$ is
\begin{equation}
\label{a12.2}
  E_{\alpha}=-\sqrt{-g}\,\,\varepsilon_{\alpha\beta\gamma}
({v^{\beta}}/{c}) B^{\gamma}-E_{\alpha}^{ext}~.
\end{equation}
The
$3-$velocities $v^{\alpha}$ are given by
\begin{equation}
\label{a7}
v^{\alpha}=\frac{c\,dx^{\alpha}}{\sqrt{h}\,dx^0}~,\quad
dx^0=c\,dt~,
\end{equation}
\begin{equation}
\label{a8}
u^{\alpha}=\frac{v^{\alpha}}{c}\left(1-\frac{v^2}{c^2}\right)^{-1/2}~,
\end{equation}
where
$$
u^0=\frac{1}{\sqrt{h}}
\left(1-\frac{v^2}{c^2}\right)^{-1/2}~,\quad
v^2=\gamma_{\alpha\beta}v^{\alpha}v^{\alpha}=
v_{\alpha}v^{\alpha}~.
$$

Substituting the expression (\ref{a12.2}) into
equation (\ref{a13a}), gives the
following equation for the magnetic field $B^{\alpha}$
\begin{equation}
\label{a15}
\frac{\partial(\sqrt{\gamma}B^{\alpha})}{\partial t}=
\frac{\partial}{\partial x^{\beta}} [\sqrt{-g}(B^{\beta} v^{\alpha}
  - B^{\alpha} v^{\beta})]+\varepsilon^{\alpha\beta\gamma}
  \frac{\partial E_{\gamma}^{ext}}{\partial x^{\beta}}\,c~.
\end{equation}
\begin{equation}
\label{a15a}
\frac{\partial}{\partial
x^{\alpha}}(\sqrt{\gamma}B^{\alpha})=0~.
\end{equation}
For the Schwarzschild metric (\ref{a2}) $(x^0, x^1, x^2, x^3)=
(ct, r, \theta, \phi)$, it follows from (\ref{a3}), (\ref{a7}),
(\ref{a8})
that $v^{\alpha}=(v^r, 0, 0)$ and
\begin{equation}
\label{a9}
v^r=c\sqrt{1-\frac{r_g}{r}}u^r
=-c\sqrt{1-\frac{r_g}{r}}\sqrt{\frac{r_g}{r}}~,
\end{equation}
where $
{v^2}/{c^2}={r_g}/{r}.$
The value of $h$, the determinant $g$ of the
$4-$metric tensor $g_{ik}$, and the determinant $\gamma$
of the metric tensor $\gamma_{\alpha\beta}$
in the Schwarzschild metric are
\begin{equation}
\label{a10}
  h=1-\frac{r_g}{r},\,\,\, \sqrt{-g}=r^2 \sin\theta,\,\,\,
\sqrt{\gamma}=\left(1-\frac{r_g}{r}\right)^{-1/2}r^2
\sin\theta~.
\end{equation}

     With $B^{\alpha}=(B^r,B^{\theta},0)$ and
$E^{ext}_{\alpha}=(0,0,E_{\phi}^{ext})$, and with all quantities
independent on the azimuthal angle $\phi$,
equations (\ref{a15}) and (\ref{a15a})
give
\begin{equation}
\label{a15.1}
\frac{\partial(\sqrt{\gamma}B^r)}{\partial t}
+\sqrt{h} v^r\frac{\partial(\sqrt{\gamma}B^r)}{\partial r}=
c\,\frac{\partial E_{\phi}^{ext}}{\partial \theta}~,
\end{equation}
\begin{equation}
\label{a15.2}
\frac{\partial(\sqrt{-g}v^rB^{\theta})}{\partial t}
+\sqrt{h} v^r\frac{\partial(\sqrt{-g}v^rB^{\theta})}{\partial r}=
-c\,v^r\sqrt{h}\frac{\partial E_{\phi}^{ext}}{\partial r}~,
\end{equation}
\begin{equation}
\label{a14}
\frac{\partial}{\partial
r}(\sqrt{\gamma}B^r)+
\frac{\partial}{\partial
\theta}(\sqrt{\gamma}B^{\theta})=0~.
\end{equation}
Equations (\ref{a15.1}) and (\ref{a15.2}) with known right-hand-sides
are solved using the characteristic method.

   The integrals of the characteristic
equations can be written as
\begin{equation}
\label{a17}
t-\int_{r_0}^r\,\frac{dr}{\sqrt{h} v^r}=C_1~,
\end{equation}
\begin{equation}
\label{a17.1}
\sqrt{\gamma}B^r-c\int_{r_0}^r \frac{\partial E_{\phi}^{ext}}{\partial \theta}
\frac{dr}{v^r\sqrt{h}}=C_2~,
\end{equation}
\begin{equation}
\label{a17.2}
\sqrt{-g}v^rB^{\theta}+c\int_{r_0}^r
\frac{\partial E_{\phi}^{ext}}{\partial r} dr=C_3.
\end{equation}
The constants $C_i$ are determined by the initial conditions.
  For the present
problem these are
$$
B^r=B^{\theta}=0,\,\,\, r=r_0\,\,\,
{\rm at}\,\,\, t=0~,
$$
which implies $C_i=0$.
In the general case the constants $C_2,\,\,C_3$
are determined by the initial values of $B^r,\,\, B^{\theta}$,
which should satisfy zero divergence condition (\ref{a14}).
We may in general take $C_1=0$, fixing the reference frame
$r=r_0$ at $t=0$.
    With account of equations (\ref{a9}) and (\ref{a10}),
equations (\ref{a17.1}) and (\ref{a17.2})
can be written as
\begin{equation}
\label{a17.3}
B^r=-\frac{\sqrt{r-r_g}}{r^{5/2}\sqrt{r_g}\sin{\theta}}
\frac{\partial}{\partial\theta}\biggl[\int_{r_0}^r E_{\phi}^{ext}
\frac{r^{3/2}dr}{r-r_g}\biggr]~,
\end{equation}
\begin{equation}
\label{a17.4}
B^{\theta}=\frac{1}{r\sqrt{r-r_g}\sqrt{r_g}\sin{\theta}}
[ E_{\phi}^{ext}(r,\theta)- E_{\phi}^{ext}(r_0,\theta)]~.
\end{equation}

    Integration of equation (\ref{a17}) with $C_1=0$ gives
a relation between
$t$, $r$ and $r_0$ in the form
\begin{equation}
\label{a20}
\frac{ct}{r_g}+\frac{2}{3}x^{3/2}+2x^{1/2}+\ln{\frac{\sqrt{x}-1}{\sqrt{x}+1}}
\end{equation}
$$
=\frac{2}{3}x_0^{3/2}+2x_0^{1/2}+\ln{\frac{\sqrt{x_0}-1}{\sqrt{x_0}+1}}~,
$$
where
$
x={r}{/r_g},$ and  $x_0={r_0}/{r_g}$ (Bisnovatyi-Kogan and
Ruzmaikin, 1974).

    Consider now the Poynting-Robertson EMF,
$E^{ext}_{\phi}(r,\theta)$.
   First we show
that $E^{ext}_{\phi}$ must  tend
to zero as $r\rightarrow r_g$
at least as fast as ($r-r_g$) or faster
in order to
avoid   singularity at
$r_g$.
    This means that in the
comoving reference system with
metric
\begin{equation}
\label{a12.4}
ds^2=
c^2 d\tau^2-\frac{r_g}{r}
d\rho^2-r^2(d\theta^2+\sin^2{\theta} d\phi^2)~,
\end{equation}
   there is
no singularity at the black hole horizon.
The connection between Schwarzschild and comoving coordinates
$(\tau,~ \rho)$ (the angle coordinates $\theta$ and $\phi$ are the same)
is
\begin{equation}
\label{a12.5}
c\tau=ct+r_g\biggl[2\sqrt{x}+\ln{\frac{\sqrt{x}-1}{\sqrt{x}+1}}\biggr],
\end{equation}
$$\rho=ct+ r_g\biggl[\frac{2}{3}x^{3/2}+2x^{1/2}+
\ln{\frac{\sqrt{x}-1}{\sqrt{x}+1}}\biggr]
$$
We can now connect the magnetic field in a comoving system
$\acute{B}^{\alpha}=(\acute B^{\rho},\acute B^{\theta}, 0)$
with the field in the
Schwarszschild system in terms of
Schwarszschild variables ($r,t$) as
\begin{equation}
\label{a12.6}
\acute B^{\rho}=\frac{r}{\sqrt{r-r_g}\sqrt{r_g}}B^r~,
\end{equation}
$$
\acute B^{\theta}=\frac{\sqrt{r-r_g}}{\sqrt{r}}B^{\theta}
+\sqrt{\frac{r_g}{r}}\frac{E^{ext}_{\phi}}{r(r-r_g)\sin\theta}~.
$$
It follows from equations  (\ref{a17.3}), (\ref{a17.4}), and  (\ref{a12.6})
that there is no singularity in the comoving frame if
$E^{ext}_{\phi}$ tends to zero   as $(r-r_g)$
or faster as $r\rightarrow r_g$.
    The metric tensor in this system (\ref{a12.4}) is
regular on the horizon, so with finite $\acute B^{\alpha}$
all $4-$invariants (for
example, $F_{ik}F^{ik}$) are also regular there.
    In fact, we
can obtain the dependence of $E^{ext}_{\phi}$  from equation
(\ref{a1}), taking into account that
the covariant component $E_{\phi}^{ext}$ in
equations (\ref{a15.1}) and (\ref{a15.2}) is
connected with the physical component $E_{\phi}^{(ph)}$
from equation (\ref{a1}) as
\begin{equation}
\label{a15.3}
  E_{\phi}^{ext}=\sqrt{\gamma_{\phi\phi}} E_{\phi}^{(ph)}
  =r\sin{\theta} E_{\phi}^{(ph)}~.
\end{equation}
   The luminosity $L$ seen by a distant observer viewing
collapsing matter with  constant comoving luminosity $L_0$ is
\begin{equation}
\label{a15.4}
L=L_0 \left(1-\frac{r_g}{r}\right)^4~,
\end{equation}
(Zeldovich \& Novikov 1971).
Thus we have from equations (\ref{a1}), (\ref{a1.1}), (\ref{a15.3}), and
(\ref{a15.4})
\begin{equation}
\label{a16}
  E_{\phi}^{ext}=
D r^{-3/2}\left(1-\frac{r_g}{r}\right)^4\sin^2{\theta}~,
\end{equation}
where
\begin{equation}
\label{a16.1}
D=\frac{L_0\sigma_T A \sqrt{GM}}{4\pi c^2|e|}~.
\end{equation}
   Equation (\ref{a16}) is of course simplified but
the dependence allows an estimate to be made
of the magnetic field generation by the PR effect
close to a black hole.
   It is necessary that $E_{\phi}^{(ph)}$ vanish
sufficiently rapidly   on the horizon in order
to avoid a physical singularity, but the exact
dependence is not important.

   Substituting equation (\ref{a16}) into (\ref{a15.1}) and
(\ref{a15.2}) gives
$$
B^r=\frac{2D}{\sqrt{r_g}}\frac{\sqrt{r-r_g}}{r^{5/2}}
\biggl[\ln{\frac{r_0}{r}}-\frac{1}{3}\left(\frac{r_g}{r}\right)^3
$$
\begin{equation}
\label{a18}
+\frac{3}{2}\left(\frac{r_g}{r}\right)^2-3\frac{r_g}{r}
+\frac{1}{3}\left(\frac{r_g}{r_0}\right)^3
-\frac{3}{2}\left(\frac{r_g}{r_0}\right)^2+3\frac{r_g}{r_0}
\biggr]\,\cos{\theta}~,
\end{equation}
$$
B^{\theta}=\frac{D}{\sqrt{r_g} r \sqrt{r-r_g}}
\biggl[\frac{1}{r^{3/2}}\left(1-\frac{r_g}{r}\right)^4 $$
\begin{equation}
\label{a19}
-\frac{1}{r_0^{3/2}}
\left(1-\frac{r_g}{r_0}\right)^4\biggr]\sin{\theta}~.
\end{equation}
There are several limiting cases where  expressions for $B^r$ and
$B^{\theta}$ can be written in a simpler form.

1. {\bf Non-Relativistic, Newtonian Regime:}
Here, $r,\,\,r_0\, \gg r_g$, and from equation (\ref{a20})
we have
\begin{equation}
\label{a21}
x_0=x\left(1+\frac{3}{2x^{3/2}}\frac{ct}{r_g}\right)^{2/3}~,
\end{equation}
and from equations (\ref{a18}) and (\ref{a19}),
\begin{equation}
\label{a22}
B^r=\frac{4D\cos{\theta}}{3r^2\sqrt{r_g}}
\ln{\left(1+\frac{3}{2x^{3/2}}\frac{ct}{r_g}\right)}
\end{equation}
\begin{equation}
\label{a23}
B^{\theta}=\frac{D\sin{\theta}}{ r^3\sqrt{r_g}}
\biggl[1-\left(1+\frac{3}{2x^{3/2}}\frac{ct}{r_g}\right)^{-1}\biggr]~.
\end{equation}
  We see here that for large $t$ the physical value $rB^{\theta}$
tends to a finite limit of the order of equation (\ref{eq26}), while
$B^r$ grows but only logarithmically.
    During the accretion time to a massive black hole
this logarithm does not exceed  $\sim 25$.
     Thus the  magnetic field is larger by
a factor $\sim 25$ than the  estimations of
equations (\ref{eq26}) and (\ref{eq27}).
    Still, the magnetic field is enormously less than
the value found by CK.

2. {\bf Vicinity of the Gravitational Radius}:
Here, $(x-1) \ll 1$ and we have from equation (\ref{a20})
\begin{equation}
\label{a23.1}
x-1=4\frac{x_0-1}{(\sqrt{x_0}+1)^2}\exp\left({-\frac{ct}{r_g}
-\frac{8}{3}+\frac{2}{3}x_0^{3/2}+2x_0^{1/2}}\right)~.
\end{equation}

     For  matter
initially  situated in the
vicinity of the horizon the Lagrangian
coordinate  $x_0-1\ll 1$.
In this case we have
\begin{equation}
\label{a25}
\ln{\frac{x_0-1}{x-1}}=\frac{ct}{r_g}~.
\end{equation}
   From equation (\ref{a18}),
\begin{equation}
\label{a26}
B^r=\frac{D\cos\theta}{2\,r_g^{5/2}}(x-1)^{9/2}\,[\exp(4ct/r_g)-1]~.
\end{equation}
  From equation (\ref{a19}),
\begin{equation}
\label{a27}
B^{\theta}=-\frac{D\sin{\theta}}{r_g^{7/2}}(x-1)^{7/2}
[\exp(4ct/r_g)-1]~.
\end{equation}
  These relations are valid also at large $t$
while $(x-1)e^{ct/r_g} \ll 1$.

    For matter with an intermediate Lagrangian coordinate,
$(x_0-1) \sim 1$, the vicinity of the horizon $r_g$ is reached only
at very large $t$, so that
\begin{equation}
\label{a27.1}
B^r={2D}\cos{\theta}\frac{\sqrt{x-1}}{r_g^{5/2}}
\biggl(\ln{x_0}-\frac{11}{6}
+\frac{1}{3x_0^3}
-\frac{3}{2x_0^2}+\frac{3}{x_0}
\biggr)~,
\end{equation}
\begin{equation}
\label{a27.2}
B^{\theta}=-\frac{D\sin{\theta}}{ r_g^{7/2}\sqrt{x-1}}
\frac{(x_0-1)^4}{x_0^{11/2}}~.
\end{equation}
We see that in the vicinity of the horizon $(x-1) \ll 1$ for
matter with an intermediate $x_0$,  $(x_0-1) \sim 1$, the component
$B^{\theta}$ in the Schwarzschild coordinates diverges.
   However, this is a natural coordinate singularity which
means only that a Schwarzschild observer cannot exist physically
in this region.
   Consequently,
the magnetic field in Schwarzschild coordinates has
observable consequences
  only in the region $x \gg 1$, $r \gg r_g$.

    Matter at a very large Lagrangian radius reaches the
vicinity of the gravitational radius at very late times.
Consider a case with
\begin{equation}
\label{a27.3}
x_0-1 \gg x,\quad x-1 \ll 1.
\end{equation}
Here, for very large $ct/r_g \gg |\ln(x-1)|$ we have
$x_0=(3{ct}/{2r_g})^{2/3}$ and
\begin{equation}
\label{a27.4}
B^r=\frac{4D\cos\theta}{3r_g^{5/2}}(x-1)^{1/2}\,
\ln{\left(\frac{3ct}{2r_g}\right)},
\end{equation}
If we assume, in addition that
$x_0 >> (x-1)^{-8/3}$, we obtain
\begin{equation}
\label{a27.5}
B^{\theta}=\frac{D\sin{\theta}}{r_g^{7/2}}(x-1)^{7/2}.
\end{equation}
 From a comparison of the asymptotic relations
  (\ref{a26}) - (\ref{a27.4}), we see that for $r$ approaching
$r_g$, the Schwarzschild component $B^r$
grows exponentially with time very close to
the horizon while remaining zero
at the horizon.
  For larger $r$ it grows logarithmically which is the
dependence in the Newtonian domain.
    The Schwarzschild component $B^\theta$
  has the mentioned singularity
on the horizon,
  which is unobservable because
a physically realizable observer cannot
measure it.
    However, at
  any fixed value of $r$ close enough to $r_g$, the temporal
behavior of the $B^{\theta}$  can be obtained from equations
(\ref{a27}), (\ref{a27.2}), and (\ref{a27.5}).
   In this region
$B^{\theta}$ starts to grow
exponentially with time, but at long times
it tends to a constant value.
This component rapidly decreases with increasing $r$.

      It is of interest to determine the magnetic
field ``seen'' by an
observer located far  from the black
hole who measures the field near the gravitational radius by
means of the cyclotron radiation coming from this region.
   The
observer determines the field strength by measuring
the cyclotron frequency of the radiation.
    The frequency of the emitted radiation is determined
by the comoving
field strength, that is, by $\acute B^{\rho}$ and
$\acute B^{\theta}$ from (\ref{a12.6}) evaluated in the vicinity
of $r_g$.
     We now obtain estimates of these fields.
   It follows from equation
(\ref{a23.1}) that matter with  Lagrangian coordinate $x_0$
approaches the horizon as $t \rightarrow \infty$.
    The
coordinate $x_0$  parametrizes the horizon points and
the radial coordinate $x$ parametrizes points of the initial
hypersphere $t=0$.
   From equations (\ref{a12.6}),  (\ref{a18}),
(\ref{a19}), (\ref{a27.1}), and (\ref{a27.2})
we obtain the  comoving fields in this region,
\begin{equation}
\label{a28}
\acute B^{\rho}=\frac{{2D}\cos{\theta}}{r_g^{5/2}}
\biggl(\ln{x_0}-\frac{11}{6}
+\frac{1}{3x_0^3}
-\frac{3}{2x_0^2}+\frac{3}{x_0}~,
\biggr)
\end{equation}
\begin{equation}
\label{a29}
\acute B^{\theta}=-\frac{D\sin{\theta}}{ r_g^{7/2}}
\frac{(x_0-1)^4}{x_0^{11/2}}~.
\end{equation}
It is easy to verify, that for any value of $r_0$, $r_g <r_0 <
\infty$, we have
\begin{equation}
\label{a30}
\acute B^{\rho}<\frac{{2D}|\cos{\theta}|}{r_g^{5/2}}\ln{x_0}~.
\end{equation}
For accretion to a stellar mass or massive black hole the
logarithmic factor  does  not exceed $\sim 25$, and the
value of $\acute B^{\rho}$ remains of the same  order
of magnitude as the
Schwarzschild component $B^r$ in the Newtonian region,
equation (\ref{a22})
at large times, formally extrapolated to $r \sim r_g$.
    The poloidal component $\acute
B^{\theta}$, given by equation (\ref{a29}), is equal to zero at
$r_0=r_g$, and $r_0=\infty$, and has a maximum at $r=11
r_g/3$. Thus
\begin{equation}
\label{a31}
\acute B^{\theta}< |\acute B^{\theta}|_{max}=\frac{\lambda D\sin{\theta}}{
r_g^{7/2}}~,
\end{equation}
where $\lambda=(3^3 8^8 11^{-11})^{1/2} \approx 0.04$.
Therefore, the
possible values of $\acute B^{\theta}$ are less than the
Newtonian value of the Schwarzschild component $ B^{\theta}$
(\ref{a23}) at large time near $r_g$.

    The proper cyclotron frequency $\omega_0$ of radiation emitted
at  $(r,t)$, but measured with respect to the comoving
time is
\begin{equation}
\label{a32}
\omega_0=\frac{|e|}{mc} \biggl[\sqrt{{r_g\over r}}(\acute B^{\rho})^2+
r^2(\acute B^{\theta})^2\biggr]^{1/2}~,
\end{equation}
where $m$ is the electron rest mass.
    Using equations (\ref{a30}) and (\ref{a31}),
we obtain an estimate for the upper limit on this
frequency,
\begin{equation}
\label{a33}
(\omega_0)_{max}\sim \frac{D|e|\sqrt{2}}{mc r_g^{5/2}}
\sqrt{\ln^2{x_0} +\frac{\lambda^2}{4}}~.
\end{equation}
The frequency  measured by a distant observer $\omega$ is
is related to $\omega_0$ as
\begin{equation}
\label{a34}
\omega=\omega_0 \sqrt{h} \frac{\sqrt{1-v^2/c^2}}{1-v\cos\psi/c}~,
\end{equation}
where $v/c=\sqrt{r_g/r}$ and $\psi$ is
the angle between $v^{\alpha}$  (which is in the negative
$r$-direction) and the direction of the photon trajectory in
Schwarzschild coordinates.
    For a radially emitted photon
($\psi=\pi$) we have from equation (\ref{a34}),$\omega=\omega_0
(1-\sqrt{r_g/r})$ near the horizon,
and for the tangential direction ($\psi=\pi/2$) we
have  $\omega=\omega_0 (1-{r_g/r})$.
     Due to the upper
limit on $\omega_0$, a distant observer does not see the
light emitted very  close to the horizon,
because of the very large red shift.
    Thus for a distant   observer the
relativistic region close to the horizon is unobservable.

    Thus we conclude that the magnetic field
produced by the PR effect close to the horizon of a black hole can be
safely estimated using the
Newtonian approximation given by equations (\ref{a22}) and (\ref{a23}).
The estimated magnetic fields are dynamically insignificant.

\section{Conclusion}

    We have reconsidered the battery effect in
accretion flows due to the non-potential
nature of the radiation force on the
electrons.
   We considered cases of a
geometrically-thin,
optically-thick disk where
a toroidal magnetic field is generated,
and a geometrically-thick, optically-thin
ADAF type accretion flow where a poloidal
magnetic field is generated due to the
Poynting-Robertson effect.
    For a stellar mass black hole the generated
toroidal field is estimated to be $\leq 10$ G, while the poloidal
field in an ADAF flow is $\leq 0.01$ G.
   The fields vary inversely with the black
hole mass.  In both cases the fields are dynamically
insignificant.
    The very large fields
obtained by CK resulted from assuming unrestricted linear growth
of the magnetic field.   The field grows only during the accretion
time.  A General Relativstic treatment of the Poynting-Robertson
generated magnetic field close to the horizon of a black hole
shows that that the field magnitude may be  larger by a factor
$\leq 25$ than the values obtained with a non-relativistic
treatment.
     Even though the magnetic field
due to the radiation force is weak,
it may have a role as a seed field
for an $\alpha-\omega$ dynamo (see for example
Colgate and Li 2000).

{\rm Acknowledgements} {This work was supported in part by NSF
grant AST-9320068. Also, this work was made possible in part by
Grant No. RP1-173 of the U.S. Civilian R\&D Foundation for the
Independent States of the Former Soviet Union. The work of RVEL
was also supported in part by NASA grant NAGW 2293. The work of
GSBK was partly supported by INTAS-ESA grant 99-120 and RAN
program ``Non-Stationary Phenomena in Astronomy.''}

\section*{Appendix A:\\ Equation for Poloidal
Magnetic Field Due to PR Effect}

    Here, we treat in more detail the influence
of the azimuthal radiation force $F_\phi^{PR}$
which   is {\it rotational}, and it {\it cannot}
be balanced by any axisymmetric electrostatic field.
    The radiation is mainly from the central
region of the  flow so that the
radiation flux-density is $S \approx L/(4\pi R^2)$,
with $L$ the accretion luminosity, and $R$ the
distance from the origin.
   The PR radiation force on an electron is
$F_\phi^{PR} = -S \sigma_T v_\phi/c^2$, where
$\sigma_T$ is the Thompson cross section
and $v_\phi$ is the azimuthal velocity of the
accreting matter.
     Including the radiation force, Ohm's for
the plasma is
\begin{equation}
\label{b1}
{\bf J} = \sigma_e \left( {\bf E}^{PR} +{\bf E}
+{\bf v \times B}/c \right)~,
\end{equation}
where $\sigma_e$ is the electrical
conductivity, and ${\bf E}^{PR} =
\hat{\rvecphi~}S\sigma_T v_\phi/(|e|c^2)$
is the Poynting-Robertson electric field.

    Combining Faraday's and Amp\`ere's laws and
equation (\ref{b1}) gives
\begin{equation}
\label{b2}
{d\Psi \over dt}\equiv
{\partial \Psi \over \partial t}
+{\bf v \cdot \nabla} \Psi = cr E_\phi^{PR}
+\eta_e \Delta^\star \Psi~,
\end{equation}
where $\eta_e =c^2/(4\pi \sigma_e)$ is the
magnetic diffusivity, and
$\Psi=rA_\phi$ is the flux function and
$A_\phi$ is the toroidal component of the vector
potential.
     Also, $\Delta^\star = \partial^2/\partial r^2
-(1/r)\partial/\partial r +\partial^2/\partial z^2$
in cylindrical coordinates and $\Delta^* =
\partial^2/\partial R^2 +[(1-\mu^2)/R^2]\partial^2/
\partial \mu^2$ in spherical coordinates where
$\mu=\cos\theta$.
    Note that $B_r= -(1/r)\partial \Psi/\partial z$
and $B_z = (1/r)\partial  \Psi/\partial r$
in cylindrical coordinates, while
$B_R=(R^2\sin\theta)^{-1}
\partial \Psi/\partial\theta$,
and $B_\theta =-(R\sin\theta)^{-1}\partial \Psi/\partial R
$ in spherical coordinates.
     Taking $v_\phi =(GM/R)^{1/2}g(\theta)$,
with $g(\pi/2)=1$ and $g(\theta \rightarrow
0,\pi)=0$,  we find
\begin{equation}
\label{b3}
E^{PR}_\phi={m_pc^2g\over 6^{3/2} |e|}
{L \over L_{Edd}}
{r_{in}^{3/2} \over R^{5/2}}
\approx{2gE_0\over 6^{3/2}}
{L\over L_{Edd}} \left(
{r_{in}\over R}\right)^{5/2}~,
\end{equation}
where $E_0(M)$ is given by equation (\ref{eq7}).
Equation (\ref{b2}) is equivalent to equation (8)
of CK.

      For an ADAF flow, the poloidal velocity
is $v_R = - \alpha \xi (GM/R)^{1/2}$ where $\alpha$ is the
Shakura-Sunyaev parameter and $\xi \leq 1$ is a constant (Narayan
\& Yi 1995).
     Following Contopoulos and Kazanas
we write $\eta_e ={\cal P} R |v_R|$,
where ${\cal P}$, the magnetic
Prandtl number, is the ratio of magnetic
diffusivity to viscosity.
    Measuring $R$ in units of $r_{in}$
and $t$ in units of $t_0=r_{in}^{3/2}/(\alpha \xi
\sqrt{GM})=\sqrt{6}(r_{in}/c)/(\alpha \xi)$,
   equation (\ref{b3}) becomes
\begin{equation}
\label{b4}
{\partial \Psi \over \partial t}\!\!
=\!\!{Kg(\theta)\sin\theta \over {R}^{3/2}}
+{1\over \sqrt{R}}{\partial \Psi \over \partial R}
+{\cal P}\sqrt{R}
\left({\partial^2 \Psi \over \partial R^2}
+{1-\mu^2 \over R^2}{\partial^2 \Psi\over \partial
\mu^2}
\right),
\end{equation}
where $K \equiv r_{in}^2 E_0(L/L_{Edd})/(3\alpha \xi)$.

    The time-scale of the linear growth of $\Psi$
is $\tau_m = t_0/{\cal P}$.
    For a turbulent magnetic diffusivity
where ${\cal P} ={\cal O}(1)$,
this time-scale is quite
short, $\sim t_0 \approx 7.3\times 10^{-4}{\rm s}
(M/M_\odot)(0.1/\alpha)(1/\xi)$.
    Therefore, the physically relevant
solution to equation (\ref{b4}) is the stationary
one  where the Poynting-Roberston  term $\propto K$ is
balanced by diffusion.
     This gives $K g(\theta)\sin\theta =
-{\cal P}(1-\mu^2)\partial^2 \Psi/\partial \mu^2$
so that $\Psi$ is independent of $R$.
       For example, for $g(\theta)=\sin(\theta)$,
$B_R = (K/{\cal P})\cos\theta/R^2$
or
\begin{equation}
\label{b5}
B_R^{PR} \approx  {0.6 \over \alpha}
{\cos\theta \over \xi {\cal P}}
{L \over L_{Edd} } {M \over M_\odot}
\left({r_{in} \over R}\right)^2~{\rm G}~,
\end{equation}
and $B_\theta=0$.  This estimate agrees with
equation (24).
   Equation (\ref{b5}) corresponds to a
radially outward field in the northern hemisphere
and a radially inward field in the southern hemisphere.
   The polarity of the field agrees
with the PR drag on the electrons
  in the $-\hat{\rvecphi~}$ giving
a ring current in the $+\hat{\rvecphi~}$
direction, while the
simple nature of the field results from the
approximation that $S\approx L/(4\pi R^2)$.

\end{document}